\documentclass{aip-cp}

\usepackage[numbers]{natbib}
\usepackage{rotating}
\usepackage{graphicx}

\begin{document}

\title{Positrons at JLab \\ Advancing Nuclear Science in Hall B }

\author[aff1]{Volker D. Burkert\corref{cor1}}

\affil[aff1]{Jefferson Laboratory, 12000 Jefferson Avenue, Newport News, Virginia, 23606} 
\corresp[cor1]{Corresponding author: burkert@jlab.org}

\maketitle

\begin{abstract}
In this talk I address two high impact physics programs that require the use of polarized and unpolarized 
positron beams in addition to using electron beams of the same energy.  First, I address what will be gained from 
using positron beams in addition to electron beams in the extraction of the Compton 
Form Factors  (CFFs) and generalized parton distributions (GPDs) from 
Deeply Virtual Compton Scattering (DVCS) on a proton target. As a second high impact science program 
 I discuss an experimental scenario using unpolarized positrons to measure elastic scattering 
 on protons in an effort to determine definitively the 2-photon exchange contributions in order to resolve a 
 longstanding discrepancy in the determination of the proton's electric and magnetic form factors.  
 \end{abstract}

\section{INTRODUCTION}
\label{intro}
The challenge of understanding nucleon electromagnetic structure still 
continues after six decades of experimental scrutiny. From the initial 
measurements of elastic form factors to the accurate determination of 
parton distributions through deep inelastic scattering, the
experiments have increased in statistical and systematic accuracy.  
During the past two decades it was realized that the parton distribution functions
represent special cases of a more general, much more powerful, way to 
characterize the structure of the nucleon, the generalized parton 
distributions (GPDs) (see~\cite{Diehl:2003ny,Belitsky:2005qn} for reviews).

\begin{figure}[h]
\includegraphics[height=0.25\textheight]{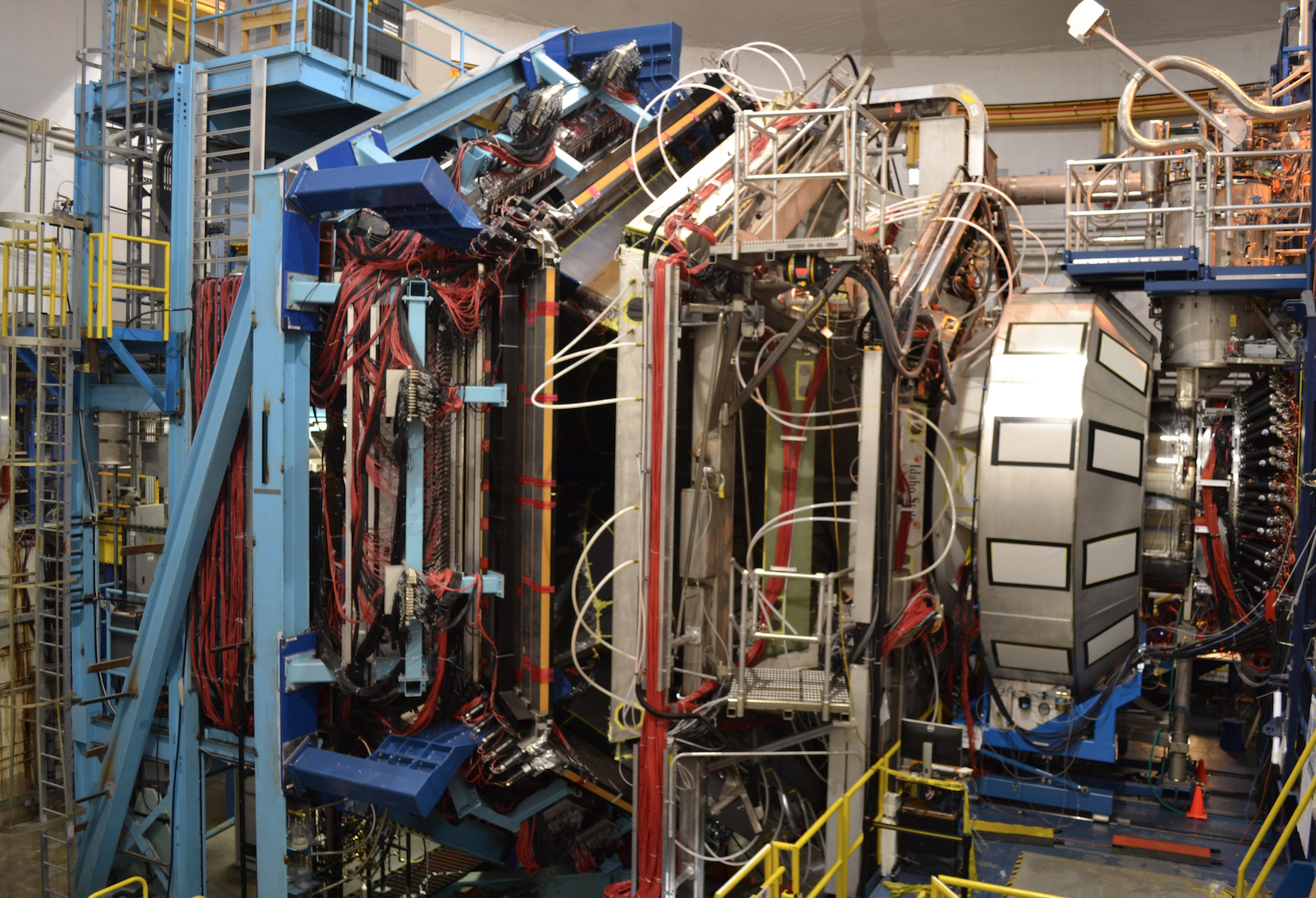}
\caption{The {\tt CLAS12} detector in Hall B. The detector was designed for inclusive, 
semi-inclusive, as well as exclusive processes such as DVCS. The construction and commissioning 
of the detector system was completed recently. {\tt CLAS12} is part of the DOE funded energy upgrade 
of the Jefferson Lab CEBAF accelerator from 
6 GeV to 12 GeV, and may play an important role in programs that make use of positron beams 
at Jefferson Lab..} 
\label{fig:clas12}
\end{figure}

 The GPDs are the Wigner quantum phase space 
distribution of quarks in the nucleon describing the 
simultaneous distribution of particles with respect to both position and 
momentum in a quantum-mechanical system.
In addition to the information about the spatial density 
and momentum density, these functions reveal the 
correlation of the spatial and momentum distributions, {\it i.e.} how the 
spatial shape of the nucleon changes when probing quarks of 
different momentum fraction of he nucleon.

The concept of GPDs has led to completely new methods of ``spatial imaging''
of the nucleon in the form of (2+1)-dimensional tomographic images, with 2 spatial
dimensions and 1 dimension in momentum~\cite{Belitsky:2003nz,Ji:2003ak,Burkardt:2002hr}. 
The second moments of GPDs are related to 
form factors that allow us to quantify how the orbital motion of quarks in the nucleon contributes to the 
nucleon spin, and how the quark masses and the forces on quarks are distributed in 
transverse space, a question of crucial importance for our understanding of 
the dynamics underlying nucleon structure and the forces leading to color confinement.   

The four leading twist GPDs $H$, $\tilde{H}$, $E$, and $\tilde{E}$, 
depend on the 3 variable $x$, $\xi$, and $t$, where $x$ is the longitudinal
momentum fraction of the struck quark, $\xi$ is the longitudinal momentum transfer
to the quark ($\xi \approx x_B/(2-x_B)$), and $t$ is the invariant 
4-momentum transfer to the proton.
The mapping of the nucleon GPDs, and a detailed understanding of the
spatial quark and gluon structure of the nucleon, have been widely 
recognized as key objectives of nuclear physics of the 
next decades. This requires a comprehensive program, combining results
of measurements of a variety of processes in electron--nucleon 
scattering with structural information obtained from theoretical studies, 
as well as with expected results from future lattice QCD simulations. The {\tt CLAS12} 
detector, shown in Fig.~\ref{fig:clas12}, has recently been completed and has begun the 
experimental science program in the 12 GeV era Jefferson Lab.  
\begin{figure}[b]
\includegraphics[height=0.2\textheight]{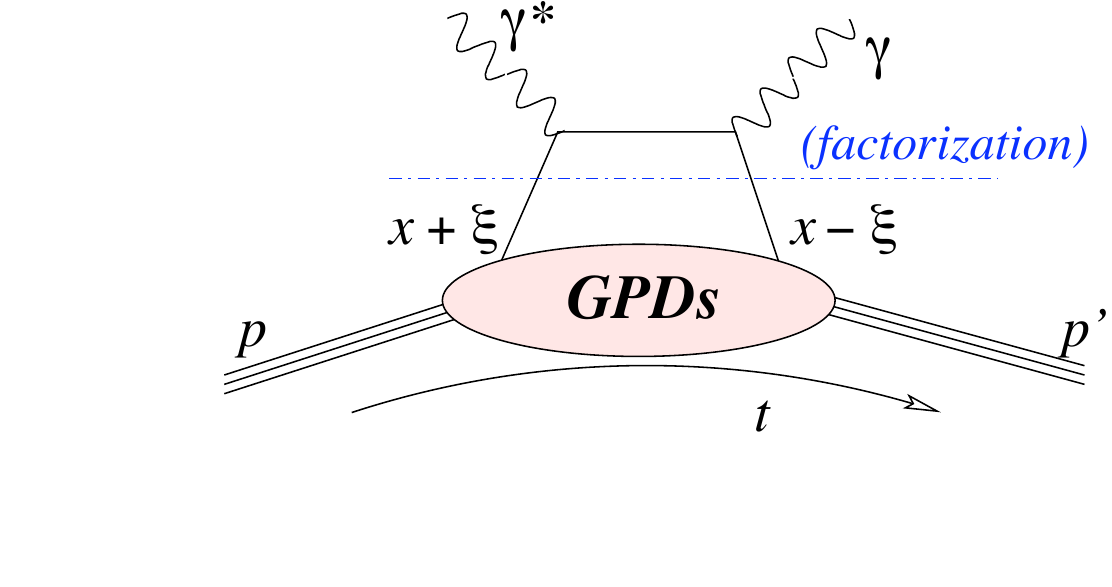}
\caption{Leading order contributions to the production of high energy single photons from protons. 
The DVCS handbag diagram contains the information on the unknown GPDs.}
\label{fig:handbag}
\end{figure}

\section {ACCESSING GPD IN DVCS}
\label{gpds} 
The most direct way of accessing GPDs at lower energies is through the measurement 
of Deeply Virtual Compton Scattering (DVCS) in a kinematical domain where the 
so-called handbag diagram shown in Fig. ~\ref{fig:handbag} makes the dominant contributions. 
However, in DVCS as in other deeply virtual reactions, the GPDs do not appear directly in the 
cross section, but in convolution integrals, e.g. 
\begin{equation}
 \int_{-1}^{+1}{{H^q(x,\xi,t)dx}\over {x - \xi + i\epsilon}} = \int_{-1}^{+1}{{H^q(x,\xi,t)dx}\over {x - \xi }} + i\pi H^q(\xi,\xi,t)~,
 \end{equation}
 where the first term on the r.h.s. corresponds to the real part and the second term to the imaginary part of the 
scattering amplitude. The superscript $q$ indicates that GPDs depend on the quark flavor. From the 
above expression it is obvious that GPDs, in general, can not be accessed directly in measurements. However, 
in some kinematical regions the Bethe-Heitler (BH) process where high energy photons are 
emitted from the incoming and scattered electrons, can be important. Since the BH amplitude is  
purely real, the interference with the DVCS amplitude isolates the imaginary part of the DVCS 
amplitude. The interference of the two 
processes offers the unique possibility to determine GPDs directly at the singular kinematics $x=\xi$. 
At other kinematical regions 
a deconvolution of the cross section is required to determine the kinematic dependencies of the 
GPDs. It is therefore important to obtain all possible independent information that will aid in extracting 
information on GPDs. The interference 
terms for polarized beam $I_{LU}$, longitudinally polarized target $I_{UL}$, transversely (in scattering plane) 
polarized target $I_{UT}$, and perpendicularly (to scattering plane) polarized target $I_{UP}$ are given 
by the expressions: 
\begin{equation}
I_{LU} \sim  \sqrt{\tau^\prime} [F_1 H + \xi (F_1+F_2) \tilde{H} + \tau F_2 E] 
\end{equation}
\begin{equation}
I_{UL} \sim  \sqrt{\tau^\prime} [F_1 \tilde{H} + \xi (F_1 + F_2) H + (\tau F_2 - \xi F_1)\xi \tilde{E}]
\end{equation}
\begin{equation}
I_{UP} \sim {\tau}[F_2 H - F_1 E + \xi (F_1 + F_2)\xi \tilde{E}
\end{equation}
\begin{equation}
I_{UT} \sim  {\tau}[F_2 \tilde{H} + \xi (F_1 + F_2) E - (F_1+ \xi F_2) \xi \tilde{E}]
\end{equation}
\noindent
where $\tau = -t/4M^2$, $\tau^\prime = (t_0 - t)/4M^2$. By measuring all 4 combinations of interference 
terms one can separate all 4 leading twist GPDs at the specific kinematics $x=\xi$. Experiments at 
JLab using 4 to 6 GeV electron beams have
been carried out with polarized beams ~\cite{Jo:2015ema,Gavalian:2008aa,Girod:2007aa,Camacho:2006qlk,Stepanyan:2001sm} and with longitudinal
target~\cite{Pisano:2015iqa,Seder:2014cdc,Chen:2006na}, showing the feasibility of such measurements at relatively low beam energies, and
their sensitivity to the GPDs. Techniques of how to extract GPDs from existing DVCS data and what has been learned about GPDs can be found in ~\cite{Guidal:2013rya,Kumericki:2012yz}.  In the following sections we discuss what information may be gained by employing both electron and positron beams in deeply virtual photon production.

\subsection{Differential cross section for polarized electrons and positrons (leptons)}
\label{crs-leptons}
The structure of the differential cross section for polarized beam and unpolarized target is given by:
\begin{equation}
\sigma_{\vec{e}p\rightarrow e\gamma p} = \sigma_{BH} + e_\ell \sigma_{INT} + P_\ell e_\ell \tilde\sigma_{INT} + \sigma_{VCS} + P_\ell \tilde\sigma_{VCS} 
\label{diffcrs}
\end{equation}
\noindent 
where $\sigma$ is even in azimuthal angle $\phi$, and $\tilde\sigma$ is odd in $\phi$. The interference terms
 $\sigma_{INT} \sim \rm{Re} {\it A}_{\gamma^*N\rightarrow \gamma N} $ and   
 $\tilde\sigma_{INT} \sim \rm{Im} {\it A}_{\gamma^*N\rightarrow \gamma N} $ are 
 the real and imaginary parts, respectively of the Compton amplitude. Using polarized 
 electrons the combination $-\tilde\sigma_{INT} + \tilde\sigma_{VCS}$ can be determined 
 by taking the difference of the beam helicities. The electron-positron charge difference for 
 unpolarized beams determines $\sigma_{INT}$. For fixed beam polarization and taking the 
 electron-positron difference one can extract the combination $P_\ell\tilde\sigma_{INT} + \sigma_{INT}$. 
If only a polarized electron beam is available one can separate $\tilde\sigma_{INT}$ 
from $\tilde\sigma_{VCS}$ using the Rosenbluth technique~\cite{Rosenbluth:1950yq}. This requires measurements at 
two significantly different beam energies which reduces the kinematical coverage that can be achieved 
with this method. With polarized electrons and polarized positrons 
both $\sigma_{INT}$ can be determined and  $\tilde\sigma_{INT}$ can be 
separated from $\tilde\sigma_{VCS}$ in the full kinematic range available at the 
maximum beam energy.   
\begin{figure}[htb]
  \includegraphics[height=.24\textheight]{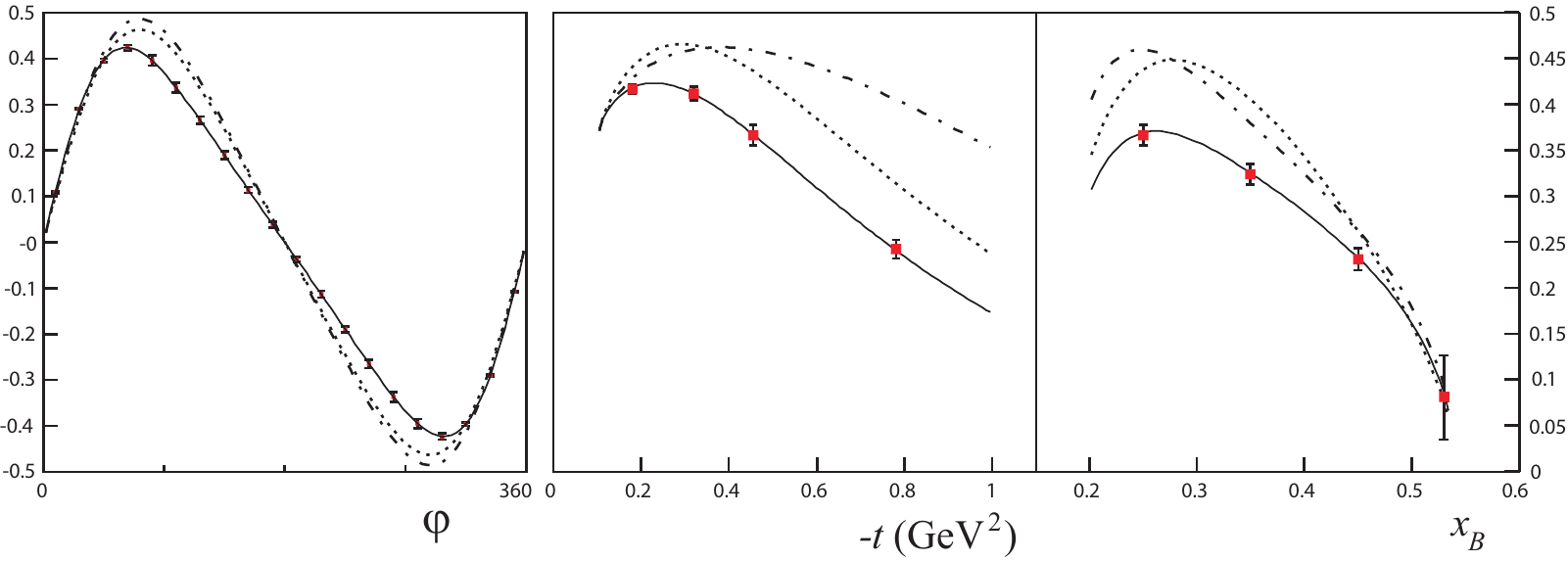}
\caption{The beam spin asymmetry showing the DVCS-BH interference for 11 GeV beam 
energy~\cite{e12-06-119}. Left panel: $x=0.2$, 
$Q^2=3.3$GeV$^2$, $-t=0.45$GeV$^2$. Middle and right panels:  $\phi=90^{\circ}$, 
other parameters same as in left panel. Many other bins will be measured simultaneously. 
The curves represent various parameterizations within the VGG model~\cite{Vanderhaeghen:1999xj}. 
Projected uncertainties are statistical.}
\label{fig:dvcs_alu_12gev}
\end{figure}

\subsection{Differential cross section for polarized proton target}
\label{crs-protons}
The structure of the differential cross section for polarized beam and polarized target 
contains the polarized beam term 
of the previous section and an additional term related to the target polarization~\cite{Belitsky:2001ns,Diehl:2005pc}:

\begin{equation}
\sigma_{\vec{e}\vec{p}\rightarrow e\gamma p} = \sigma_{\vec{e}p\rightarrow e\gamma p} +
T[P_\ell\Delta\sigma_{BH} + e_\ell \Delta\tilde\sigma_{INT} + P_\ell e_\ell\Delta\sigma_{INT} + 
\Delta\tilde\sigma_{VCS} + P_\ell\Delta\sigma_{VCS}]
\end{equation}
where the target polarization $T$ can be longitudinal or transverse. If only unpolarized electrons are available, 
the combination 
$-\Delta\tilde\sigma_{INT} + \Delta\tilde\sigma_{VCS}$ can be measured from the differences in the 
target polarizations. If unpolarized electrons and unpolarized positrons are available the combination 
$T\Delta\tilde\sigma_{INT} + \sigma_{INT}$ can be determined at fixed target polarization. 
With both polarized electron and 
polarized positron beams, the combination $T\Delta\tilde\sigma_{INT} + TP_\ell\Delta\sigma_{INT} 
+ P_\ell\tilde\sigma_{INT} + \sigma_{INT}$ can be measured at fixed target polarization. Availability 
of both polarized electron and polarized positron beams thus allows the separation of all contributing terms. 
\begin{figure}[tb]
  \includegraphics[height=.40\textheight]{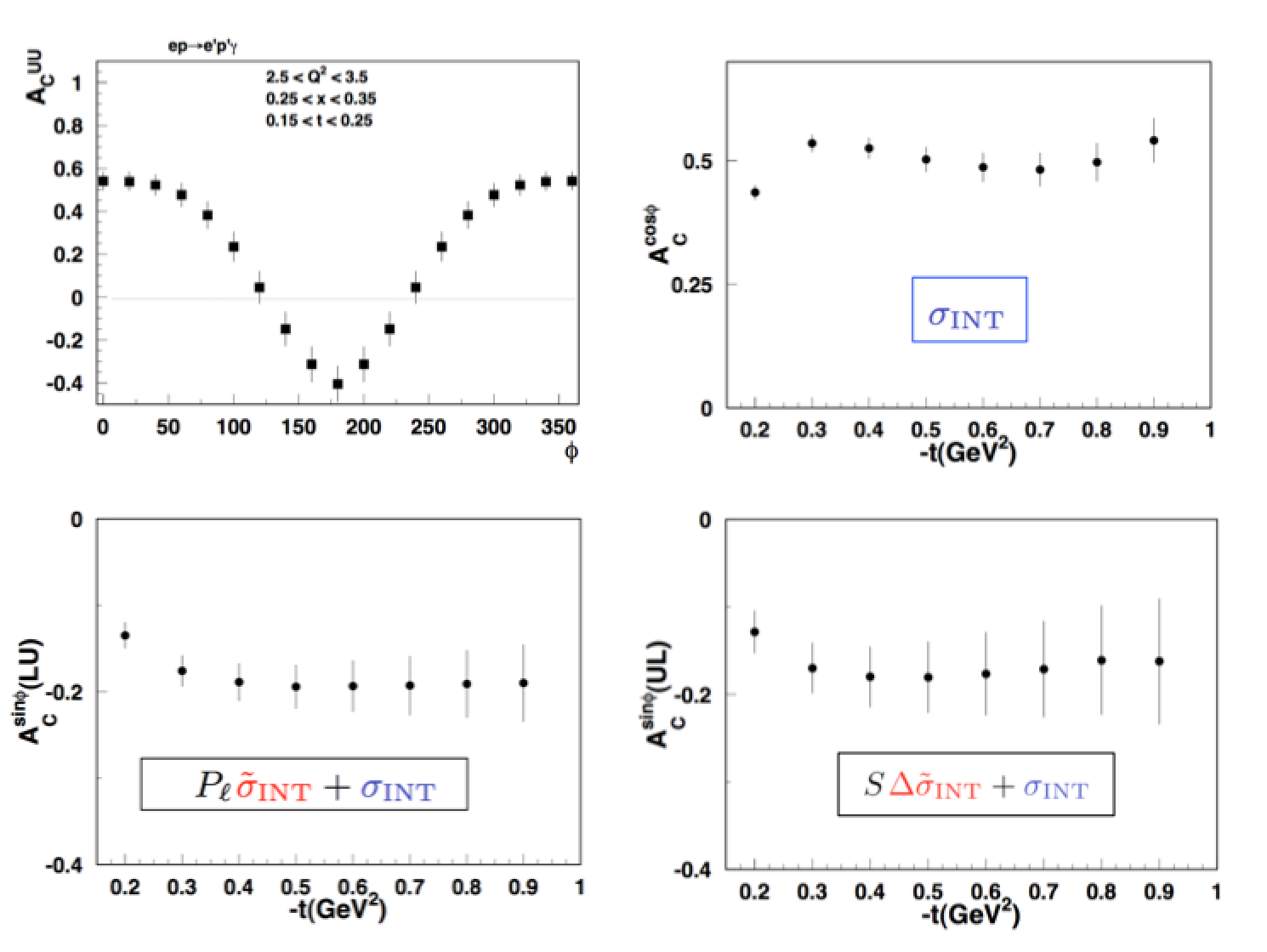}
\caption{Electron-positron DVCS charge asymmetries: Top-left: Azimuthal dependence of the charge 
asymmetry for positron and electron beam at 11 GeV beam. Top-right: Moment in $\cos(\phi)$ of the 
charge asymmetry versus momentum transfer $t$ to the proton.  
Bottom-left: Charge asymmetries for polarized electron and positron beams 
at fixed polarization (LU). Bottom right:  Charge asymmetry for longitudinally polarized protons 
at fixed polarization (UL). 
 The error bars are estimated for a 
1000 hrs run with positron beam and luminosity $L = 2\times 10^{34}$~cm$^{-2}$sec$^{-1}$ at a beam 
polarization $P=0.6$.
Electron luminosity $L = 10 \times10^{34}$~cm$^{-2}$sec$^{-1}$, and electron beam polarization $P = 0.8$. 
The error bars are statistical for a single bin in $Q^2$, $x$, and $t$ as shown in the top-left panel. Other 
bins are measured simultaneously.}
\label{fig:cross_section}
\end{figure}
If only polarized electron beams are available a Rosenbluth separation with different beam energies can
separate the term $\Delta\tilde\sigma_{INT}$ from $\Delta\tilde\sigma_{VCS}$, again in a much more limited kinematical
range and with likely larger systematic uncertainties. 
The important interference term $\Delta\sigma_{INT}$ can only be determined using the combination of polarized electron 
and polarized positron beams.  

\subsection{Estimates of charge asymmetries for different lepton charges}
For quantitative estimates of the charge differences in the cross sections we use the acceptance and 
luminosity achievable with {\tt CLAS12} as basis for measuring the process 
$ep \rightarrow e\gamma p$ at different beam and target conditions. A 10 cm long liquid hydrogen is
assumed with an electron current of 40nA, corresponding to an operating luminosity of 
$10^{35}$cm$^{-2}$sec$^{-1}$.   For the positron beam a 5 times lower beam current of 8nA is 
assumed. In either case 1000 hours of beam time is used for the rate projections. 
For quantitative estimates of the cross sections the dual model~\cite{Guzey:2008ys,Guzey:2006xi} is 
used. It incorporates parameterizations of the GPDs 
$H$ and $E$.  As shown in Fig.~\ref{fig:cross_section}, effects coming from the charge 
asymmetry can be large. In case of unpolarized beam and unpolarized target the cross section
for electron scattering has only a small dependence on azimuthal angle $\phi$, while the corresponding
positron cross section has a large $\phi$ modulation.  The difference is directly related to the 
term $\sigma_{INT}$ in equation (\ref{diffcrs}). 

\subsection{Experimental Setup for DVCS Experiments}
\label{dvcs}
\begin{figure}[h]
  \centerline{\includegraphics[height=180pt,width=350pt]{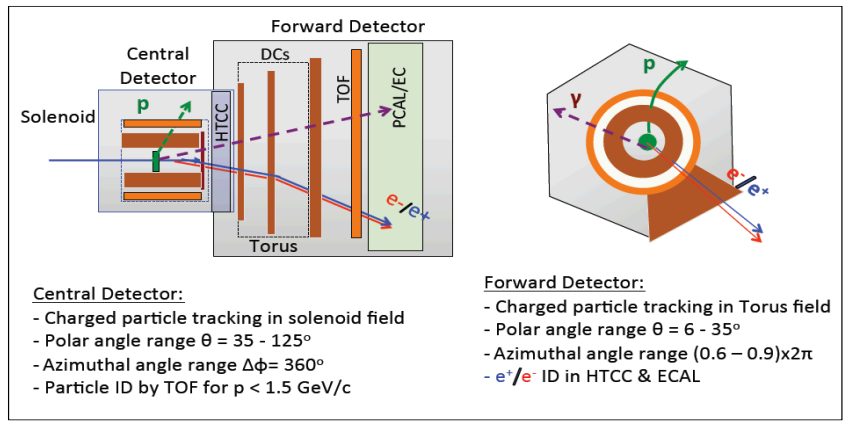}}
  \caption{CLAS12 configuration for the two electron and positron experiments (generic).  The central detector will detect
  the protons, and the bending in teh magnetic solenoid field will be identical for the same kinematics. The electron and the 
  positron, as well as the high-energy DVCS photon will be detected in the forward detector part. The electron and positron will be
  deflected in the Torus magnetic field in the same way as the Torus field direction will be opposite in the two experiments.
  The deflection in $\phi$ due to the solenoid fringe field will be of same magnitude $\Delta\phi$ but opposite in direction.
  The systematic of this shift can be controlled by doing the same experiment with opposite solenoid field directions that 
  would result in the sign change of the $\Delta\phi$. }
  \label{dvcs_experiment}
\end{figure}

Figure~\ref{dvcs_experiment} shows generically how the electron-proton and the positron-proton DVCS experiments could be 
configured. Electrons and positrons would be detected in the forward detection system of CLAS12. However, for the positron 
run the Torus magnet would have the reversed polarity so that positron trajectories would look identical to the electron trajectories 
in the electron-proton experiment, and limit systematic effects in acceptances. The recoil proton in both cases would  be detected 
in the Central Detector at the same solenoid magnet polarity, also eliminating most systematic effects in the acceptances. 
However, there is a remaining systematic difference in the two configuration, as the forward scattered electron/positron would 
experience different transverse field components in the solenoid, which will cause the opposite azimuthal motion in $\phi$ 
in the forward detector. A good understanding of the acceptances in both cases is therefore important. The high-energy photon 
is, of course, not affected by the magnetic field configuration.        

In the next section we discuss a possible solution to the, so-far, not conclusive experimental studies of two-photon effects 
in elastic electron-proton scattering and their effect on the ratio of electric to magnetic form factors $G_E/G_M$ 
versus $Q^2$. 

 \section{2-PHOTON EFFECTS IN ELASTIC SCATTERING OFF PROTONS} 
 \label{2-photon-effects}

In the electromagnetic physics community it is well known that two experimental approaches, the Rosenbluth separation 
and the beam polarization transfer approach results in conflicting values for the $G_E/G_M$ ratio when plotted as a function 
of $Q^2$. The results of the different experimental methods are compiled in Fig.~\ref{GeGm}. The trends of the two data sets 
are inconsistent with each other, although there is a large spread in the Rosenbluth data samples. The latter seem to be more consistent with near $Q^2$-independent behavior, 
while the polarization data have a strong downward behavior with $Q^2$. Furthermore, the uncertainties in the former are
 much larger 
and within the individual data sets there seem to be discrepancies as well. 
The difference of the two methods may been attributed to 2-photon exchange effects, which are expected to be 
much more important in the cross section subtraction method than in the polarization transfer method.

\subsection{Recent efforts to quantify 2-photon exchange contributions}
\begin{figure}[t]
  \centerline{\includegraphics[height=160pt,width=320pt]{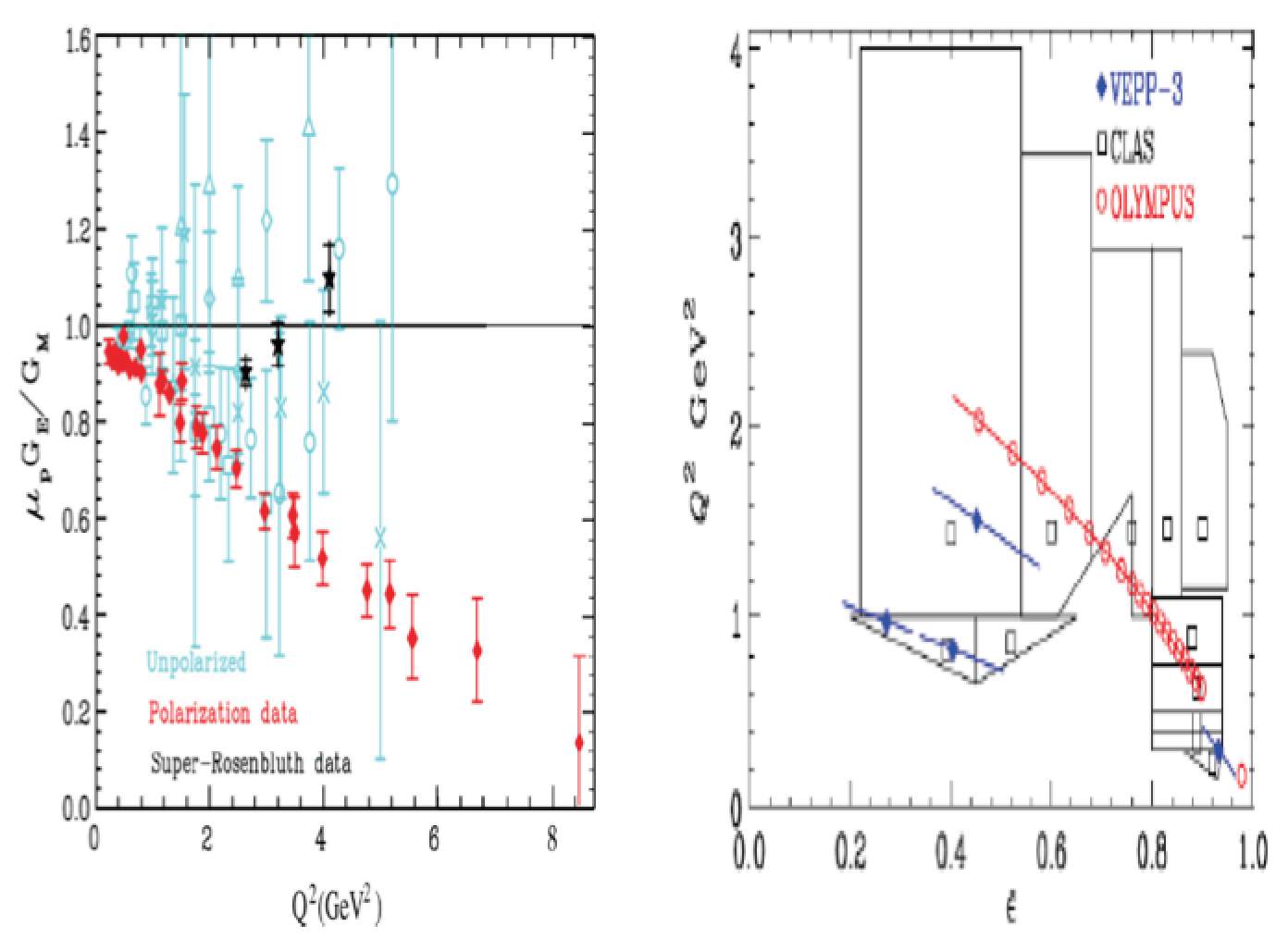}}
  \caption{Left panel: The ratio of the proton electric and magnetic form factors $G_E/G_M$. The cyan markers are results of experiments 
  based on the Rosenbluth method. The red markers are from JLab Hall A experiments. Right panel: Kinematics covered by the 
  three recent experiments to measure the 2-photon exchange contribution to the elastic ep cross section.  
  Both figures are taken from a recent review article~\cite{Afanasev:2017gsk}.}
  \label{GeGm}
\end{figure}

It is obviously important to resolve the discrepancy with experiments that have sensitivity to 2-photon contributions. The most 
straightforward process to evaluate 2-photon contribution is the measurement of the ratio of elastic $e^+p/e^-p$ scattering, 
which in leading order is given by the expression:  $R_{2\gamma} = 1 - 2\delta_{\gamma\gamma}$. 
Several experiments have recently been carried out to measure the 2-photon exchange contribution in elastic scattering: the 
VEPP-3 experiment at Novosibirsk~\cite{Rachek:2014fam}, the CLAS experiment at 
Jefferson Lab~\cite{Rimal:2016toz,Adikaram:2014ykv}, and the Olympus experiment at DESY~\cite{Henderson:2016dea}.
The kinematic reach of each experiment is shown in the right panel of Fig.~\ref{GeGm}. The kinematic coverage is much smaller 
in these experiments $Q^2 < 2$~GeV$^2$, and $\epsilon > 0.5$, where the 2-photon effects are expected to be small, and systematics
of the measurements must be extremely well controlled.  The combined evaluation of all three experiments led the authors of 
the review article Ref.~\cite{Afanasev:2017gsk} to the conclusion that the results of the experiments are inconsistent with the $\delta_{\gamma\gamma} = 0$ hypothesis at 99.5\% confidence. At the same time, they state that {\it "the results of these experiments are by no means definitive"}, and  {\it "there is a clear need for similar experiments at larger $Q^2$ and at  $\epsilon < 0.5$."}  
\subsection{Conclusions From Previous Experiments}

In the following I discuss a possible experiment with CLAS12 at Jefferson Lab that may be able to remedy the shortcomings of the 
previous measurements. What are these shortcomings? 

\begin{figure}[t]
\includegraphics[height=180pt,width=320pt]{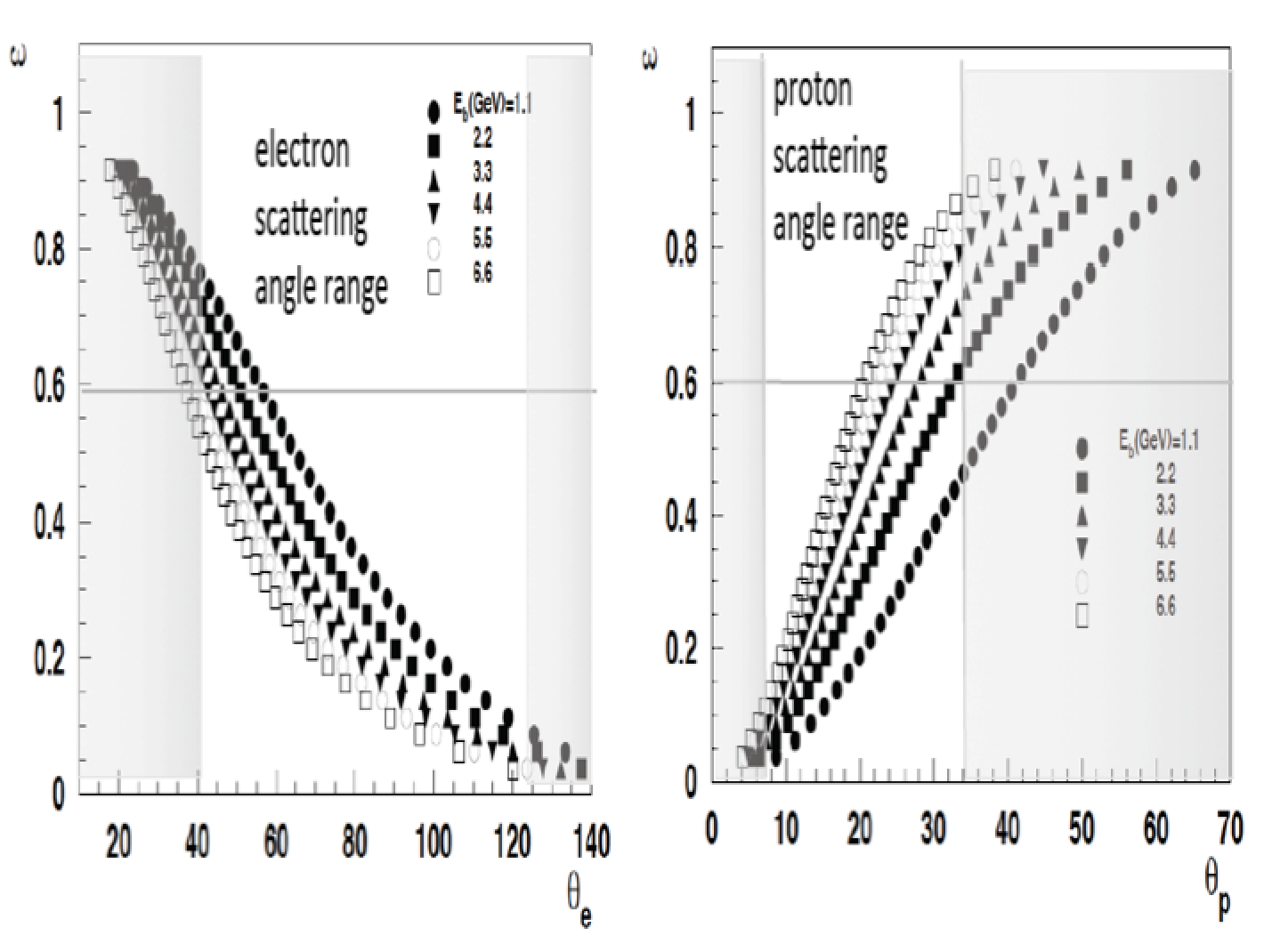}
\caption{Polar angle and $\epsilon$ coverage for electron detection (left) and for proton detection (right). }
\label{angle_reach}
\end{figure}

\begin{itemize}  
\item Kinematics coverage in $Q^2$ and in $\epsilon$ are mostly where 2-photon effects are expected to be small
\item  Systematic uncertainties are marginal in some cases
\item  Higher $Q^2$ and small $\epsilon$ corresponding to high energy and large 
electron scattering angles were out of reach
\end{itemize}
Can we do better with a setup using the modified CLAS12 detector? To address this question we begin with the 
close to ideal kinematic coverage that this setup provides. Figure~\ref{angle_reach} shows the angle coverage for both the electron (left) and for the 
proton (right). There is a one-to-one correlation between the electron scattering angle and the proton recoil angle. For the 
kinematics of interest, say $\epsilon < 0.6$ and $Q^2 > 2$~GeV$^2$ for the chosen beam energies from 2.2 to 6.6 GeV, 
nearly all of the electron scattering angles fall into a polar angle range from $40^\circ$  to $125^\circ$, and corresponding to 
the proton polar angle range from $8^\circ$ to $35^\circ$. While these kinematics are most suitable for accessing 
the 2-photon  exchange contributions, the setup will be able to also measure the reversed kinematics with the electrons at
forward angle and the protons at large polar angles. This is in fact the standard CLAS12 configuration of DVCS and most other 
experiments, however it will not cover the kinematics with highest sensitivity to the 2-photon exchange contributions.   

\begin{figure}
\includegraphics[height=180pt,width=330pt]{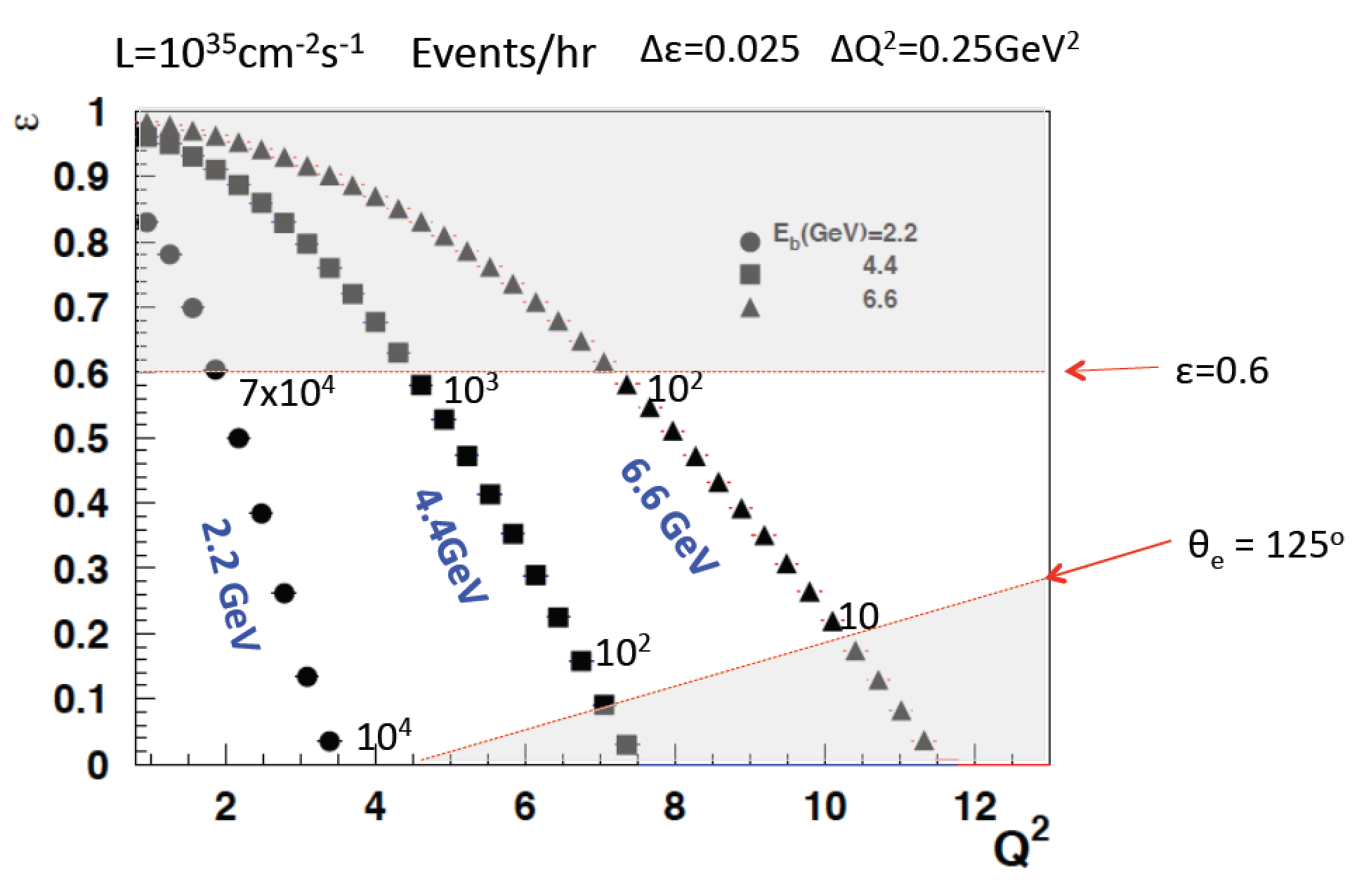}
 \caption{Estimated elastic event rates per hour for selected standard CEBAF beam energies of 2.2, 4.4, 6.6 GeV 
 in the $\epsilon$ - $Q^2$ plane.  Rates are given only for the lowest and highest $Q^2$ bin.}
 \label{Q2eps}
 \end{figure}

Figure~\ref{Q2eps} shows the expected elastic scattering rates  covering the ranges of highest interest, with 
$\epsilon < 0.6$ and $Q^2 = 2 - 10$~GeV$^2$. Sufficiently high statistics of ${\sigma_N / N}  < 1\%$ can be achieved within 10 hrs for
 the lowest energy and within 1000 hrs for the highest energy, to cover the full range in kinematics. 
 Note that all kinematic bins will be measured simultaneously at a given energy, and the shown rates are for the individual 
bins in $Q^2$ - $\epsilon$ space.


\subsection{A New Experimental Setup - Kinematic Coverage and Rate Estimates}


 \begin{figure}[h]
 \includegraphics[height=180pt,width=350pt]{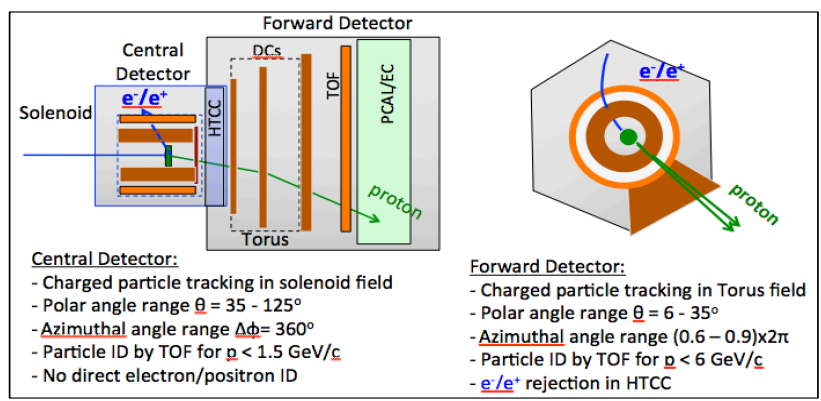}
  \caption{CLAS12 configuration for the elastic $e^-p/e^+p$ scattering experiment (generic).  The central detector will detect
  the electron/positrons, and bending in the solenoid magnetic field will be identical for the same kinematics. The proton 
  will be detected in the forward detector part.  The Torus field direction will be the same in both cases.  The deflection 
  in $\phi$ due to the solenoid fringe field will be of same in magnitude of $\Delta\phi$ but opposite in direction.
  The systematic of this shift can be controlled by doing the same experiment with opposite solenoid field directions that 
  would result in the sign change of the $\Delta\phi$.}
\label{2gamma-exp}
\end{figure}
  In order to achieve the desired reach in $Q^2$ and $\epsilon$ the {\tt CLAS12} detection system has to be used 
with reversed detection capabilities for electrons. The main modification will involve replacing the current Central Neutron 
Detector (CND) with a central electromagnetic calorimeter (CEC) . The CEC will not need very good energy or angle 
resolution (both are provided by the tracking detectors) but will be used for trigger purposes and to aid in 
electron/pion separation. The over constrained kinematics of measured scattered electrons and recoil protons should be 
sufficient to select the elastic kinematics and eliminated any background (this will have to be demonstrated by detailed 
simulations).    

\noindent
For the rate estimates and the kinematical coverage we have made a number of assumptions that are not overly stringent:
\begin{itemize}
\item Positron beam currents (unpolarized): $I_{e^+} \approx 60$~nA.  
\item Beam profile: $\sigma_x,~\sigma_y < 0.4$~mm.  
\item Polarization: not required, so phase space at the source maybe chosen for optimized yield and beam parameters.
\item Obtain the electron beam from the same source as the positrons to keep systematic under control.
\item Switching from $e^+$ to $e^-$ operation should be doable in  reasonable time frame ( $< 1$~day) to keep machine stable, 
and systematics under control. 
\item  Operate experiment with 5cm liquid H$_2$ target and luminosity of $0.8  \times 10^{35}$~cm$^2$sec$^{-1}$
\item Use the CLAS12 Central Detector for lepton ($e^+/e^-$) detection at $\Theta_l = 40 - 125^\circ$.
\item Use CLAS12 Forward Detector for proton detection at $\Theta_p = 7^\circ - 35^\circ$
 \end{itemize}   
The CLAS12 configuration suitable for this experiment is shown in Fig.~\ref{2gamma-exp}. 
\section{SUMMARY} 
Availability of a 11 GeV positron beam at JLab can significantly enhance the experimental 
program using the {\tt CLAS12} detector in Hall B~\cite{Burkert:2012rh}. I discussed two high profile 
programs that would very significantly benefit from a  high performance polarized positron source and 
accelerated beam. The first program fits well into the already developed 3D-imaging program 
with electron beams, where the imaginary part of the DVCS amplitude can be extracted.  The program 
with polarized positrons enables access to the azimuthally
even BH-DVCS interference terms that are directly related to the real part of the scattering amplitude. 
Moreover, by avoiding use of the Rosenbluth separation technique, the leading contributions to the 
cross sections may be separated in the full kinematical range available at the JLab 12 GeV upgrade.  Even 
at modest polarized positron beam currents of 8nA good statistical accuracy can be achieved for charge 
differences and charge asymmetries. For efficient use of polarized targets higher beam currents of up to
40nA are needed to compensate for the dilution factor of $\sim 0.18$ inherent in the use of currently  
available polarized proton targets based on ammonia as target material, and to allow for a more complete 
DVCS and GPD program at 12 GeV. The second program requiring positron beams is the 
measurement of the 2-photon exchange contributions in the elastic electron-proton scattering. The measurements
we outlined, if properly executed with excellent control of systematic uncertainties, should close the book   
on the discrepancies in the ratio of electric to magnetic form factors when measured with two different methods. 

In this talk we have focussed on experiments with a large acceptance detector, which may be the only option given 
the low current expected for polarized positron beams of high polarization and good beam parameters. Positron 
currents in excess of $1 \mu A$ may be required to make positron beams attractive for an experimental program
with focusing, high resolution magnetic spectrometers.



\section{Acknowledgments}
I like to thank Harut Avakian for providing me with the experimental projections for the DVCS experiment,
and the elastic ep scattering cross sections and rate calculations.   
This material is based upon work supported by the U.S. Department of Energy, Office of Science, Office of Nuclear Physics under contract DE-AC05-06OR23177.

\vfill\eject

\end{document}